\documentclass[aps,prb,twocolumn,superscriptaddress,showkeys,longbibliography,nobibnotes]{revtex4-2}
\usepackage[colorlinks=true,linkcolor=blue,urlcolor=blue,citecolor=blue,anchorcolor=blue]{hyperref}
\pdfoutput=1 
\usepackage{textcomp,gensymb}
\usepackage{graphicx}
\usepackage[T1]{fontenc}
\usepackage{txfonts}

\begin{document}
	
\title{Exploring low-temperature dynamics in triple perovskite ruthenates using nonlinear dielectric susceptibility measurements}

\author{Shruti Chakravarty}
\affiliation{Department of Physics, Indian Institute of Science Education and Research, Pune, India}

\author{Sunil Nair}
\email[Corresponding author: ]{sunil@iiserpune.ac.in}
\affiliation{Department of Physics, Indian Institute of Science Education and Research, Pune, India}

\date{\today}

\begin{abstract}
We report the nonlinear dielectric properties of three triple-perovskite ruthenates: Ba$_3$CoRu$_2$O$_9$, Ba$_3$BiRu$_2$O$_9$, and Sr$_3$CaRu$_2$O$_9$. These compounds exhibit notable correlations among their spin, charge, lattice, and polar degrees of freedom. Ba$_3$CoRu$_2$O$_9$ displays a pronounced frequency-dependent relaxation in $\chi_{2,3}$ just above the magnetoelastic transition, occurring around 100K, followed by an abrupt loss of polarization within the ordered phase. In Ba$_3$BiRu$_2$O$_9$, we encounter the possibility of multiple coexisting relaxation behaviors, indicating a complex phase strongly influenced by the spin-gap opening at 175K. Lastly, Sr$_3$CaRu$_2$O$_9$ displays anomalies with strong dispersion effects close to its magnetic transitions, pointing to a robust coupling between magnetic and dipolar orders in the system. These measurements highlight the significance of higher-order (hyper-)susceptibilities in providing profound insight into the dynamics of a system, offering information otherwise inaccessible through the linear polarization response alone.
\end{abstract}


\maketitle

\section{Introduction}
Dielectric Spectroscopy is one of the most widely used experimental techniques to understand molecular dynamics at glass transitions due to the availability of an exceptionally broad time/frequency window. In these measurements, one often measures the linear polarization response of the sample to an applied ac field by measuring the complex dielectric permittivity ($\epsilon = \epsilon' + i\epsilon''$) which is related to the dielectric susceptibility $\chi$ as $\chi$ = $\epsilon$+1. However, the full polarization response ($P$) is actually a power series written as:
\begin{equation}
	P = \epsilon_0(\chi_1E+\chi_2E^2+\chi_3E^3+....)
\end{equation}
including higher order terms with respect to the external electric field $E$ \cite{jonscher1999,miga2007}. The nonlinear components $\chi_i$ with $i>1$ are called hypersusceptibilities and contain an abundance of information which often goes unexplored. This is due to the relative difficulty in measuring them since they are many orders of magnitude weaker than the linear response. However, in many cases, these higher-order susceptibilities can provide key information otherwise inaccessible through the linear term. According to the Landau-Ginzburg-Devonshire (LGD) theory  of ferroelectric (FE) phase transitions, $\chi_i$ for $i=1,2,3$ are obtained as \cite{ikeda1987}:
\begin{equation} \label{x1}
	\chi_1 = \frac{1}{\epsilon_0[A(T-T_c)+3BP^2]}
\end{equation}
\begin{equation} \label{x2}
	\chi_2 = -3\epsilon_0BP\chi_1^3
\end{equation}
\begin{equation} \label{x3}
	\chi_3 = -(1-18\epsilon_0BP^2\chi_1)\epsilon_0^3B\chi_1^4
\end{equation}
where $\epsilon_0$ is the vacuum permittivity, $A$ is a constant, $T_c$ is the critical temperature at which the phase transition occurs, and $B$ and $C$ are smooth functions of temperature \cite{fujimoto2005}. The sign of $B$ determines the type of phase transition: $B>0$ yields a continuous (2\textsuperscript{nd} order) phase transition while $B<0$ yields a discontinuous (1\textsuperscript{st} order) transition. 

Eq. \ref{x2} shows that $\chi_2$ is proportional to the net induced polarization in the system and reflects its orientation and magnitude. Thus, in the paraelectric (PE) state, $\chi_2$ is expected to vanish and changes sign when the polarity of the applied field is reversed. On the other hand, $\chi_1$ and $\chi_3$ are related to $P^2$ and thus, have finite values even above $T_c$. In the PE phase ($P=0$), eq. \ref{x3} reduces to: $\chi_3 = -\epsilon_0^3B\chi_1^4$, which implies that $\chi_3 < 0$ for $B>0$. While in the FE phase, $\chi_3 = 8\epsilon_0^3B\chi_1^4$ which is positive for $B>0$. So, $\chi_3$ is expected to show a change of sign at a continuous FE transition. Measurement of $\chi_1$ and $\chi_3$ also allows the estimation of the scaled nonlinear susceptibility ($a_3$) given by $a_3 = \frac{-1}{\epsilon_0^3}\frac{\chi_3}{\chi_1^4}$ which in the PE state is simply equal to $B$ and in the FE state of a continuous transition, takes the value $-8B$, thus also showing a sign change. This prediction has been experimentally verified for materials exhibiting second-order FE transitions like Triglycine Sulphate (TGS, (NH$_2$CH$_2$COOH)$_3$$\cdot$H$_2$SO$_4$) \cite{miga2007TGS} and Lead Germanate (Pb$_5$Ge$_3$O$_{11}$) \cite{miga2006LGO}.  Similarly, for a discontinuous transition ($B, C < 0$), $\chi_3 > 0$ and $a_3 < 0$ is expected on both sides of $T_c$ \cite{ikeda1987, miga11}. This has again been confirmed experimentally in measurements on BaTiO$_3$, a model ferroelectric showing multiple temperature-dependent first-order transitions. This clearly illustrates that both $\chi_3$ and $a_3$ are crucial parameters which can distinguish between 1\textsuperscript{st} and 2\textsuperscript{nd} order phase transitions \cite{miga11}.  

In case of relaxor FEs which show no spontaneous symmetry breaking, the Spherical Random Bond Random Field (SRBRF) model \cite{pirc1999} assumes the constituent polar nanoregions (PNRs) to interact via a spin-glass like random-exchange coupling in presence of quenched random local electric fields. This model suggests that in presence of a dynamic external perturbation, the PNRs flip to realign with a characteristic timescale $\tau$ and yields a negative $\chi_3$ along with two extrema: one at the freezing temperature $T_f$ and another at the temperature $T_m$ at which $\chi_1$ peaks.  Consequently, $a_3$ is expected to remain positive at all temperatures. However, experiments on typical relaxor ferroelectrics like 2\% Ba-doped LGO, PbMg$_{1/3}$Nb$_{2/3}$O$_3$ (PMN) and Sr$_{0.61}$Ba$_{0.39}$Nb$_2$O$_6$ have resulted in contrasting results with positive, frequency dependent $\chi_3$ and a negative $a_3$ at all temperatures \cite{miga11}. $a_3$ is also expected to diverge strongly at $T_f$ in dipolar glasses, where $\chi_3 \propto (T-T_f)^{-\gamma}$, $\gamma=1$ being the mean field exponent \cite{pirc1994,binder1992}. This has been partially confirmed in the dipolar glass K$_{0.989}$Li$_{0.011}$TaO$_3$ (KLT1.1) where frequency-dependent peaks in $\chi_3$ have been observed followed by a sign-change and eventual drop to zero values. However, robust confirmations of proposed theories are yet elusive. 

This veritably illustrates that measurement of nonlinear susceptibilities can provide profound insight into the critical dynamics of a system at the phase transition. Hypersusceptibilities being more sensitive to dipolar orders can also help distinguish between dipolar and quadrupolar interactions. In the quantum paraelectric (PE) SrTiO$_3$, measurements of $\chi_3$ have provided experimental evidence of the onset of a coherent tunneling ground state below 33 K by directly probing the FE pair correlations between the dipole moments \cite{hemberger1995}. Phenomenological theories have also been proposed which suggest that $\chi_3$ can also probe the amplitude mode in a type-II incommensurate phase as it shows a jump with a change of sign at the incommensurate transition point \cite{iwata1998}. It is evident that the scope to explore the nonlinear response of disordered ferroelectrics is vast and the field remains a trove of open questions that are worth investigating. Thus, it is worthwhile to focus our attention on these quantities and utilize them in our quest to explore novel physics in quantum materials.

Until now, most nonlinear studies on relaxors have focused on model FE systems like A/B-site substituted perovskites \cite{dec2008PMN, bobnar2001PLZT, kleemann2014BTS, kleemann2013BTZ} or Tungsten-Bronze based compounds \cite{dec2003SBN61}. Barely any reports exist on studies done on double- or triple-perovskite systems, families of materials which have proven to be excellent hosts to a plethora of unconventional physical properties by virtue of their inherent geometrical frustration accompanied by disorder. In this study, we employ the nonlinear dielectric susceptibility in probing the correlated orders in three triple perovskite systems: Ba$_3$CoRu$_2$O$_9$, Ba$_3$BiRu$_2$O$_9$ and Sr$_3$CaRu$_2$O$_9$. 
Ba$_3$CoRu$_2$O$_9$ shows strong coupling between spin, charge and lattice degrees of freedom and simultaneously undergoes an antiferromagnetic (AFM) transition, a hexagonal-orthogonal structural transition and a semiconductor-semiconductor electronic transition (related to a change in the number of charge carriers) at $\sim$93 K. These coinciding transitions have been presented as evidence of an orbital ordering in the Ru$^{5+}$ t$_{2g}$ manifold in previous reports \cite{zhou2012} which results in distortion of the RuO$_6$ octahedra within the the Ru$_2$O$_9$ dimers and can explain the suppression of magnetic moment observed. 
Ba$_3$BiRu$_2$O$_9$ is another fascinating system, reported to show a magnetoelastic transition at $\sim$175K and is accompanied by a spin-gap opening. The material also shows anomalous features in its dielectric response, including suppression of frequency dispersion at the magnetoelastic transition and a relaxor-like behaviour above the spin-gap opening regime \cite{kumar2022}. 
Finally, Sr$_3$CaRu$_2$O$_9$ is the first ruthenate reported with a 2:1 perovskite ordering and a d$^0$ ion at the B-site. Nb$^{5+}$ and Ta$^{5+}$ compounds with similar structures are known to be good dielectrics and used in microwave applications \cite{rijssenbeek2002}. Even so, it remains relatively unexplored. In a recent work, it has been reported to show two successive magnetic transitions at $\sim$160 K and $\sim$190 K with concomitant features in the dielectric permittivity ($\epsilon(T)$) \cite{kumar2023} implying possible magnetodielectric coupling. It is evident that each of these compounds are interesting subjects to perform nonlinear dielectric investigations on.  

\begin{figure}[ht]
	\includegraphics[width=\columnwidth]{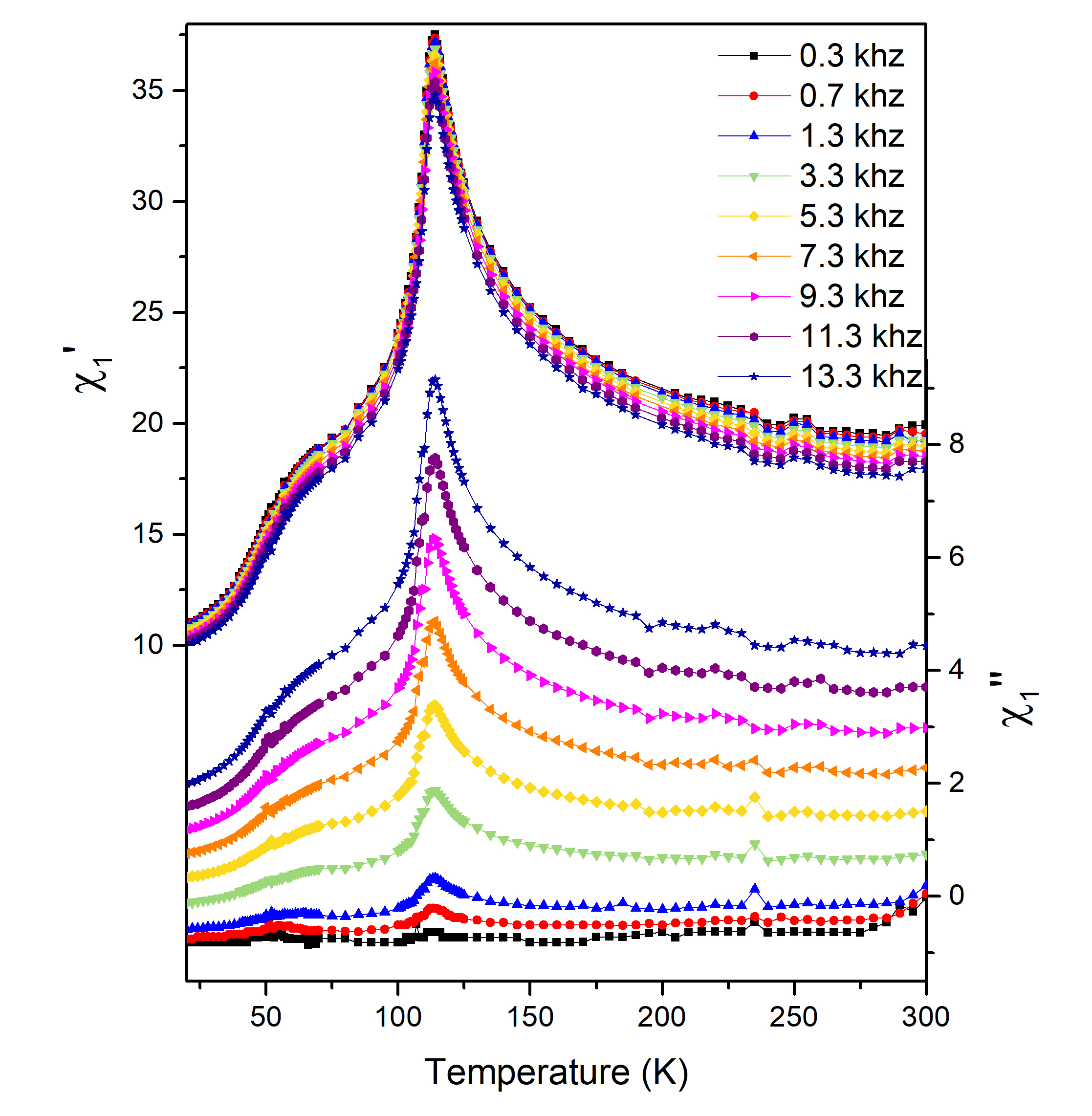}
	\caption{Real (top) and imaginary (bottom) first-order susceptibility $\chi_1$ of KDP (KH$_2$PO$_4$) as a function of temperature. Both the discontinuous FE transition and the shoulder peak associated with the domain freezing are visible, confirming the reliability of the setup}
	\label{fig1} 
\end{figure}

\section{Experimental Methods}
The materials were prepared using the conventional solid-state route using stoichiometric amounts of (Ba/Sr)CO$_3$, CaCO$_3$, RuO$_2$, Bi$_2$O$_3$ and Co$_3$O$_4$. The measurements were performed using a home-made dielectric susceptometer based on the technique reported by Miga and Kleeman \cite{miga2007} with a few modifications. Instead of using an A/D converter on a KPCI card, we use a Lock-in Amplifier (SR830) to measure the voltage signal proportional to the displacement current measured by the current pre-amplifier (SR570) which functions as a current-voltage converter in our setup. Sample response is measured by coating the pellets with conductive silver paste on the two faces similar to a parallel-plate capacitor. Using the lock-in's phase sensitive detection we are able to decouple the real and imaginary susceptibilities and independently measure them. The perturbing electric field is sourced from the lock-in amplifier itself, to eliminate any phase errors. However, since SR830 can only source and measure frequencies upto 100 kHz, we only measure upto the third harmonic (i.e. the third-order susceptibility) and limit our maximum applied frequency to 33.33 kHz. The temperature is controlled by using Lakeshore 335 temperature controller in conjunction with a Cryo Industries Closed-Cycle Refrigeration system. The temperature is varied using a cartridge heater and measured using a Lakeshore Silicon Diode sensor. The noise floor of our instrument lies around a few $\mu$Vs and the data is recorded using a fully-automated LabView program.

\section{Results and Discussion}
Initially, to calibrate the setup we performed measurements on KDP (KH$_2$PO$_4$) which is a prototypical hydrogen-bonded FE material often used for optoelectronic and piezoelectric applications. KDP shows a para- to ferro-electric (PE $\rightarrow$ FE) transition at $\sim$ 123 K, accompanied by a domain freezing at lower temperatures \cite{kuramoto1987}. Fig. \ref{fig1} shows the real and imaginary parts of the linear susceptibility measured on a circular pellet of thickness 1.1 mm made by uniaxially pressing KDP powders. The FE transition is marked by the frequency-independent discontinuity visible both in the real and imaginary susceptibility, ($\chi_1'$, $\chi_1''$) at about 118K. The broadening of the cusp is most likely due to polycrystallinity of the sample. On lowering the temperature, a hump is also clearly seen between 50-100K indicating the onset of domain freezing. Note that we do not see the peak associated with this freezing in the imaginary part of the linear susceptibility ($\chi_1$'') at frequencies above 733 Hz, however the magnitude of the loss response is strongly frequency dependent and is seen to increase with frequency even though the peak is broadened significantly. 
 
\begin{figure}[ht]
	\includegraphics[width=\columnwidth, trim=0.2cm 0.2cm 0 0]{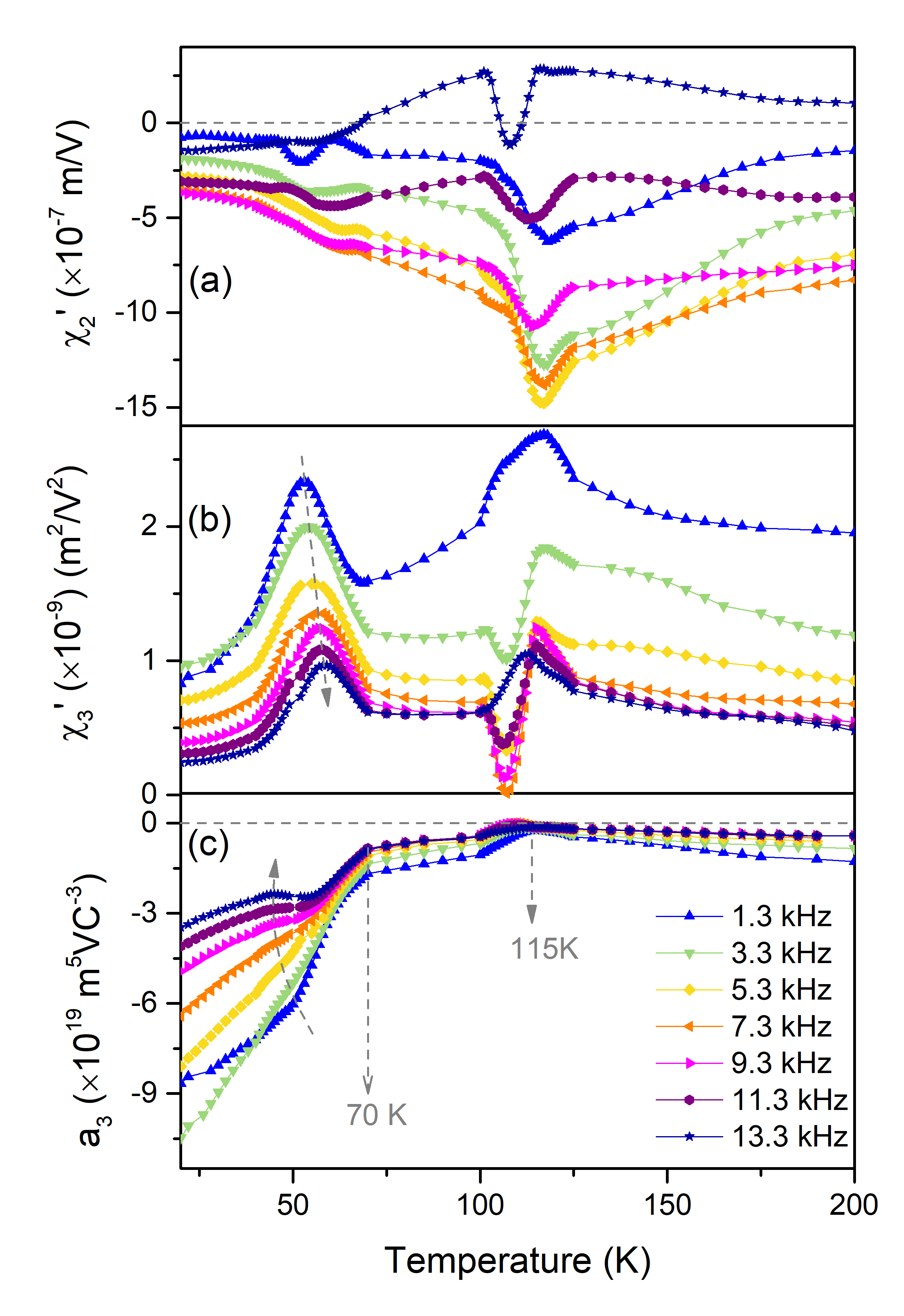}
	\caption{(a) Real second-order susceptibility $\chi_2'$ of KDP (KH$_2$PO$_4$) as a function of temperature. Polarization remains negative throughout for all frequencies except 13.3 kHz. (b) Real third-order susceptibility $\chi_3'$ of KDP (KH$_2$PO$_4$) as a function of temperature. Both the domain freezing process and the first-order PT are captured.  The latter is a broadened discontinuity possibly due to the sample being a pellet made out of polycrystalline powder. (c) Scaled nonlinear susceptibility $a_3$ of KDP (KH$_2$PO$_4$) as a function of temperature. Grey dashed arrows show position of kinks corresponding to the FE and freezing transition. Frequency dispersion is also visible below 70 K. $a_3$ remains negative throughout.}
	\label{fig2} 
\end{figure}

The real part of the second-order susceptibility ($\chi_2'$) is shown in Fig. \ref{fig2} (a). $\chi_2'$ ($= -3\epsilon_0^2BP\chi_1^3$) is a direct probe of the total polarization ($P$) within the material at a given time and hence, can be used to detect changes in $P$ occuring in the material as a function of temperature or frequency. $\chi_2'$ is expected to be non-zero only for macroscopically noncentrosymmetrical systems, otherwise the net polarization (and consequently $\chi_2'$) is expected to vanish \cite{bottcher1973}. Both the PE (\textit{I$\bar{4}$2d}) and FE (\textit{Fdd2}) phases of KDP are non-centrosymmetric \cite{jia2020}, and thus we observe finite second-order polarization throughout the measurement range with considerable frequency dependence. $\chi_2'$ remains negative throughout (except for 13.3 kHz, the highest frequency measured) the temperature range measured and shows a clear peak at the FE transition. The sign of $\chi_2'$ is generally not considered important since it depends on the net polarization orientation and can be flipped simply by changing the polarity of the electrodes \cite{miga11}. Interestingly, the magnitude of the FE transition peak keeps increasing till 5.3 kHz beyond which it starts to gradually decline again, eventually crossing over to positive values at 13.3 khz, all the while maintaining the same qualitative behaviour. This seems to reflect the presence of uncompensated domains within the FE phase which are oppositely oriented with respect to the applied field and thus start to show up only at higher frequencies where the larger domains oriented along the applied field are not able to contribute to the response as much. As the temperature continues to increase, $\chi_2'$ is diminished further to a constant finite value. The loss response, $\chi_2''$ is significantly stronger (not shown here), almost two orders of magnitude larger than the real part and is seen to mimic the fundamental response. 

Fig. \ref{fig2} (b) shows the third-order susceptibility, $\chi_3$  of KDP as a function of temperature. Both the domain freezing process and the discontinuous FE transition are captured at various frequencies. The frequency-dependence between 50-60K and positive value of $\chi_3'$ is reminiscent of relaxors like PbMg$_{1/3}$Nb$_{2/3}$O$_3$ (PMN) and Sr$_{0.61}$Ba$_{0.39}$Nb$_2$O$_6$ (SBN61) \cite{miga11} which also contradict the predictions of the SRBRF Model \cite{pirc1999}. On the other hand, $\chi_3' > 0$ at all temperatures is expected for a discontinuous FE transition by the LGD theory \cite{miga11}. Although the dispersion is clearly observable, the relaxation time obtained from the peak positions ($\tau$,$T_{peak}$) does not follow any of the Arrhenic, Vogel-Fulcher (VF) or critical slowing-down dynamics. The scaled nonlinear susceptibility, $a_3$ is also shown in Fig. \ref{fig2} (c) containing kinks corresponding to both the FE transition (at $\sim$115 K) and the subsequent domain freezing below 70 K. $a_3$, just like $\chi_3$ shows enhanced frequency dispersion at lower temperatures. These measurements confirm the reliability of our setup by confirming the first order nature of the FE transition followed by the freezing of FE domains on further cooling. Also, agreeing with the predictions of the LGD theory, for a discontinuous transition, $\chi_3$ remains $>0$ and $a_3$ remains $<0$ in the full temperature range.
 
\begin{figure}[ht]
	\includegraphics[width=\columnwidth, trim=0.2cm 0.2cm 0 0]{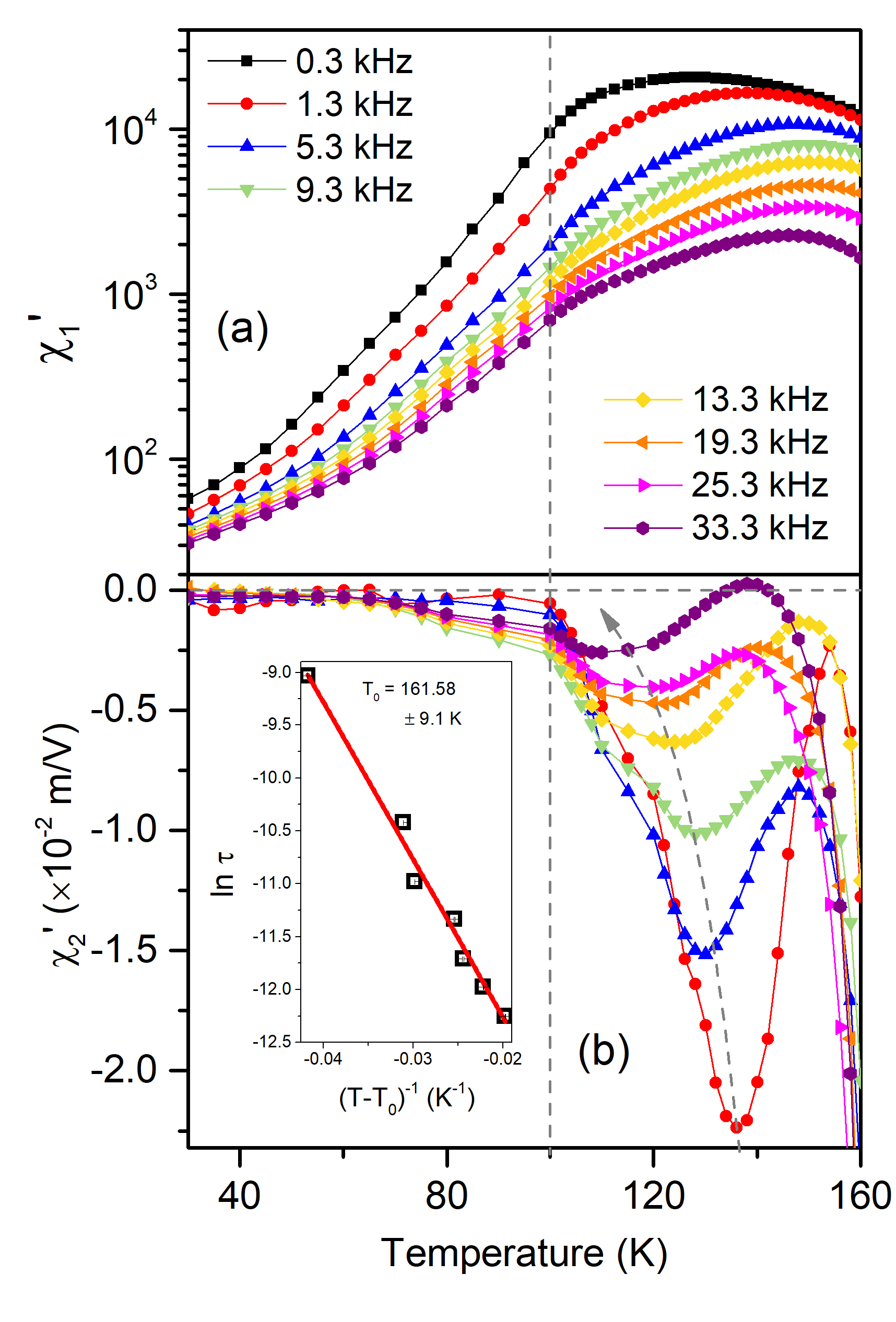}
	\caption{(a) Real first-order susceptibility $\chi_1'$ of Ba$_3$CoRu$_2$O$_9$ plotted as a function of temperature for various frequencies in log scale. The clear slope change corresponding to the magnetoelastic transition is visible and is also marked by the grey dashed line. Above 150 K, conduction effects dominate and cause $\chi_1$ to become $< 0$. (b) Real second-order susceptibility $\chi_2'$ of Ba$_3$CoRu$_2$O$_9$ plotted as a function of temperature for various frequencies in log scale. Negative polarization rises dramatically above 100 K. Peculiar dynamics observed: polarization peak shifts to lower temperatures with higher frequencies. Beyond 150 K, conduction effects dominate and negative polarization continues to sharply increase}
	\label{fig3} 
\end{figure}

Fig.\ref{fig3} shows the real part of the linear susceptibility of Ba$_3$CoRu$_2$O$_9$. We observe a clear change of slope at $\sim$100K similar to its resistivity ($\rho(T)$) behaviour \cite{zhou2012}, corresponding to the AFM ordering coinciding with an electronic transition. This indicates strong coupling of magnetic, electronic and polar orders in this material. This is not surprising since at this temperature, a strong structural distortion of the RuO$_6$ octahedra within the Ru$_2$O$_9$ dimers also results in the simultaneous occurrence of a structural transition from a higher-symmetry hexagonal phase ($P6_3/mmc$) to a lower-symmetry orthorhombic phase ($Cmcm$). The electronic transition is confirmed by the presence of two different activation energies required to describe the conduction below and above T$_N$ = 100 K \cite{zhou2012}. 

This transition is captured in greater detail in $\chi_2'$ (Fig.\ref{fig3}) in the form of a sharp rise in negative polarization in the material above 100K followed by the emergence of frequency dependent negative peaks whose relaxation dynamics is well-described by the Vogel-Fulcher (VF) relation. This relationship is unexpected as can be seen in the inset of Fig. \ref{fig3} (b) where we plot the relaxation time $\tau$ as a function of $(T-T_0)^{-1}$, where $T_0$ is the estimated "true" freezing temperature ($\approx$ 162K, here) and obtain a negative $\frac{E_a}{k_B}$ and $\tau_0 \approx$ 10$^{-7}$s.  The temperature of the extremum shifts to lower temperatures at higher frequencies, unlike commonly observed in glass- or relaxor-like systems. The reason for this peculiar behaviour is unclear.  $\chi_2'$ remains negative throughout and beyond 150K drops sharply to values upto 2 orders of magnitude higher.  The structure of Ba$_3$CoRu$_2$O$_9$ remains centrosymmetric in both phases, so $\chi_2'$ should ideally be zero both above and below T$_N$. However, since these are polycrystalline samples, there is strong possibility of accumulation of free charges at grain boundaries that contributes more and more as temperature is increased. Finite values of $\chi_2'$ have also been documented for other such materials like La$_2$NiMnO$_6$, Ca$_3$Mn$_2$O$_7$ \cite{sahlot2020}. It is also possible that the non-zero $\chi_2'$ originates from incomplete averaging to zero of the total polarization of the PNR subsystem which develop as a result of the significant bond distortion in the RuO$_6$ octahedra close to the transition. 

\begin{figure}[ht]
	\includegraphics[width=\columnwidth, trim = 0.3cm 0.2cm 0 0]{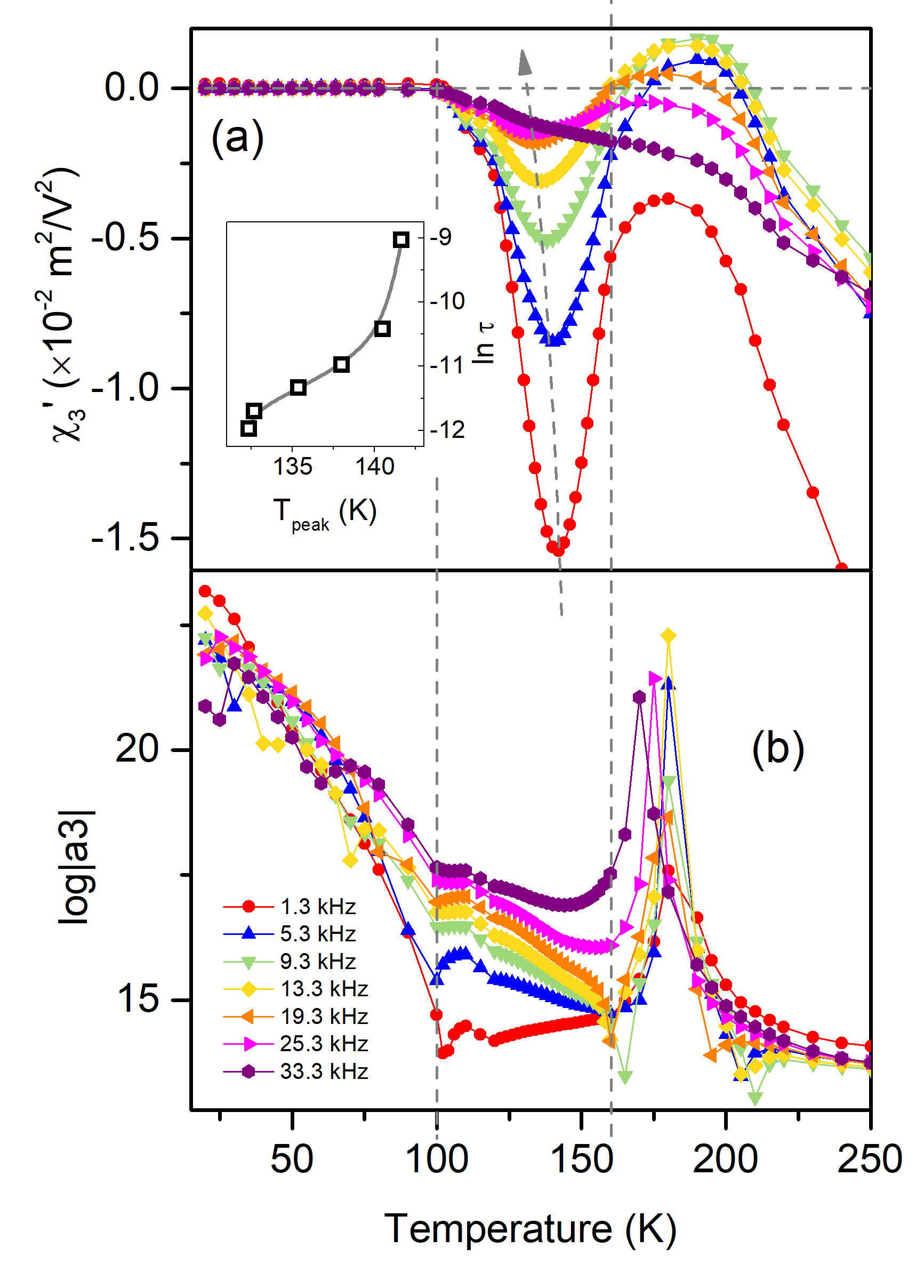}
	\caption{(a) Real third-order susceptibility $\chi_3'$ of Ba$_3$CoRu$_2$O$_9$ plotted as a function of temperature for various frequencies in log scale. $\chi_3'$ rises dramatically above 100 K on the negative scale displaying dynamics similar to $\chi_2'$. Beyond 150 K, for some freqencies $\chi_3'$ crosses over to the positive scale and peaks, but falls back to negative values above 225 K. Inset shows variation of variation of $\tau$ with the temperature of $\chi_3$ (negative) peaks. Grey line is only guide to the eye. (b) Log of absolute value of scaled nonlinear susceptibility, log $|a_3|$ of Ba$_3$CoRu$_2$O$_9$ as a function of temperature shown for frequencies 1.3, 5.3, 9.3, 13.3 19.3, 25.3 and 33.3 kHz. Gray dashed arrows mark transition temperatures.}
	\label{fig4} 
\end{figure}

We document similar behaviour in $\chi_3$ (Fig. \ref{fig4}) which shows massive rise in negative polarization above $T_N$ followed by frequency-dependent peaks between 130-150 K. The peaks again shift to lower temperatures with increasing frequency, similar to $\chi_2'$, however in this case, the variation does not agree with relaxation models like Arrhenius, VF Law or the mean field critical-slowing down (inset of Fig.\ref{fig4} (a)). The negative peak is strongly suppressed as applied frequency is increased until it is washed out completely for 33.3 kHz.  For some frequencies, $\chi_3'$ shows a crossover to positive values followed by a peak before dropping back down to negative values above $\sim$210K. $\chi_3$ remaining negative for $T>T_N$ for most of the frequencies is similar to the expectation of the SRBRF model for a relaxor FE system. It is possible that the negative peaks observed at $\sim$150 K correspond to the peak in $\chi_1$, and before $T_f$ is reached, the magnetostructural transition takes over and the local polarization is lost. The temperature variation of the scaled nonlinear susceptibility $a_3$ is also shown in Fig. \ref{fig4} (b) in the form of log$|a_3|$ vs $T$. The two discontinuities at $T_N$ and 160 K immediately stand out, the latter is due to the crossover of $\chi_1'$ to negative values, due to the overwhelming contribution of enhanced conduction at high temperatures. This is expected since the resistivity of this semiconducting material at room temperature is very small ($\sim$0.07$\ohm$-cm \cite{zhou2012}). In the intermediate temperatures, $a_3$ stays positive, in agreement with the predictions of the SRBRF model. 

\begin{figure}[ht]
	\includegraphics[width=\columnwidth, trim=0.2cm 0.3cm 0 0]{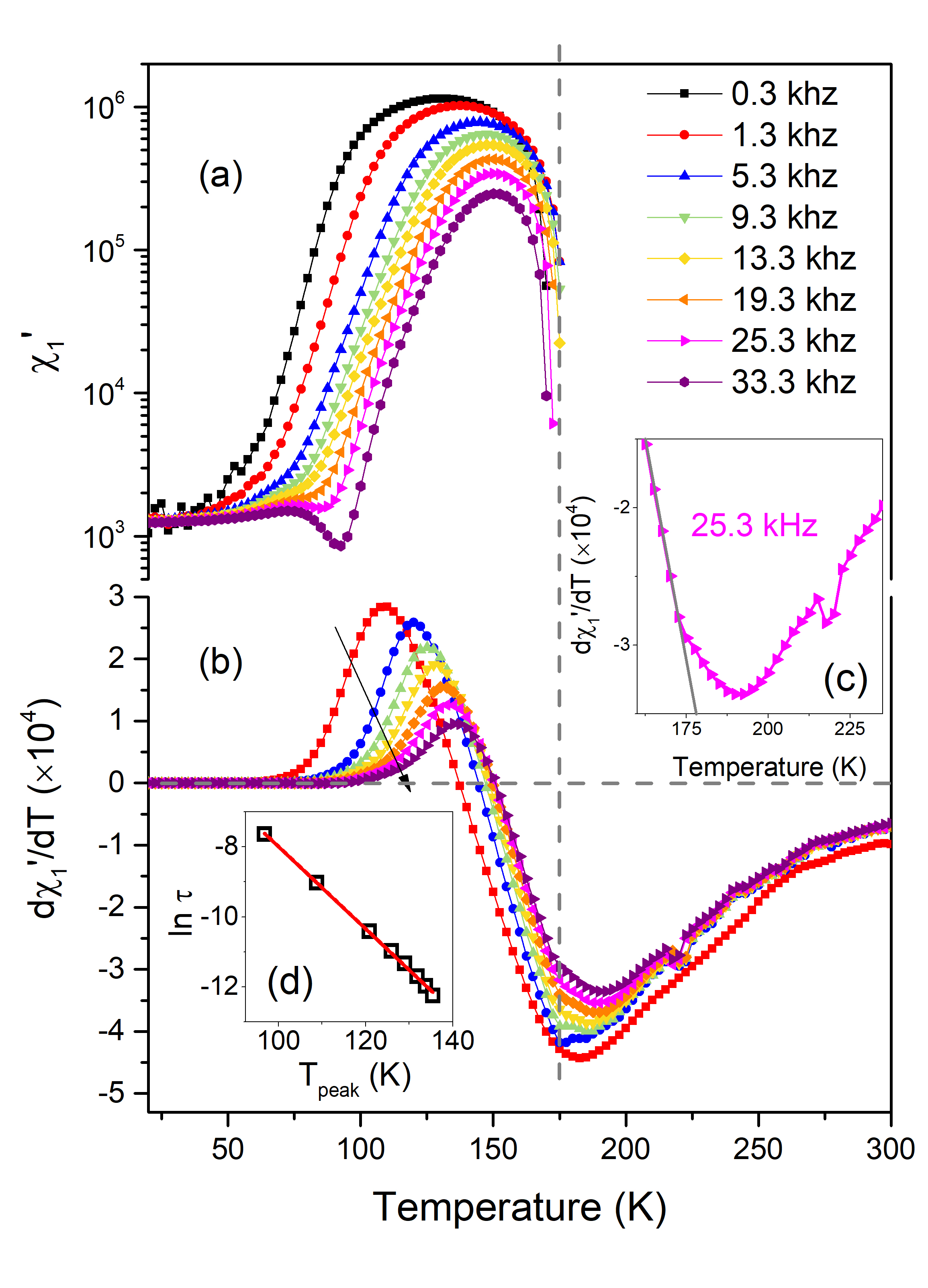}
	\caption{(a) Real first-order susceptibility $\chi_1'$ of Ba$_3$BiRu$_2$O$_9$ plotted as a function of temperature for various frequencies in log scale. $\chi_1'$ starts to nosedive to negative values close to 175 K under the influence of enhanced conduction. Feature below 100 K is most pronounced at 33.3 kHz, is suppressed at lower frequencies. (b) Derivative of real first-order susceptibility d$\chi_1'$/dT of Ba$_3$BiRu$_2$O$_9$ plotted as a function of temperature for various frequencies in order to distinguish the spin-gap opening transition from the overtake of leakage current. (c) Zoom-in of d$\chi_1'$/dT at $f=$ 33.33 kHz showing the clear slope change corresponding to the spin-gap transition. (d) frequency ($\tau = 1/(2\pi f)$) dependence of the peak corresponding to the feature below 100 K with a linear fit.}
	\label{fig5} 
\end{figure}

Fig. \ref{fig5} (a) shows the linear dielectric susceptibility $\chi_1'$ of Ba$_3$BiRu$_2$O$_9$  plotted on a log scale. The curves display dispersion across a wide temperature range along with change in susceptibility spanning decades. In this case too, like Ba$_3$CoRu$_2$O$_9$, the conductivity of the sample overwhelms the linear response and drives $\chi_1'$ to negative values above 175 K.  This is different from previous reports \cite{kumar2022} where permittivity $\epsilon'$ (proportional to $\chi_1'$) remains positive upto room temperature. One of the possible reasons for this could be defect formation due to different synthesis/sintering conditions, since sample sintering strongly affects grain boundary formation, which in turn affects bulk conduction and charge polarization response. This is a problem, since the magnetoelastic transition and the spin-gap opening are also expected to occur in this temperature regime. In an attempt to resolve this, we plot the derivative of $\chi_1'$ with respect to $T$ (Fig. \ref{fig5} (b)) which is able to pick up the transition point in spite of its proximity to the leakage current dominated region. This is shown in Fig. \ref{fig5} (c), where zooming into the transition region unveils a very subtle slope change at 175 K signifying the spin-gap opening transition, while the upturn close to 200 K marks the leakage current taking over. There is significant frequency dependence between 100-150 K corresponding to a divergence-like feature observed in $\chi_1'$ which is seen to be suppressed at lower frequencies. Hints of this feature were seen in a previous report in linear dielectric measurements, but were attributed to artifacts \cite{kumar2022}. $\tau$ vs $T_{peak}$ inferred from the relaxation peaks show a linear behaviour with a negative slope (Fig. 5 (d)) which does not corroborate with any of the usual relaxation models (Arrhenius, VF, critical slowing down etc.).

\begin{figure}[ht]
	\includegraphics[width=\columnwidth, trim=0.2cm 0.2cm 0 0]{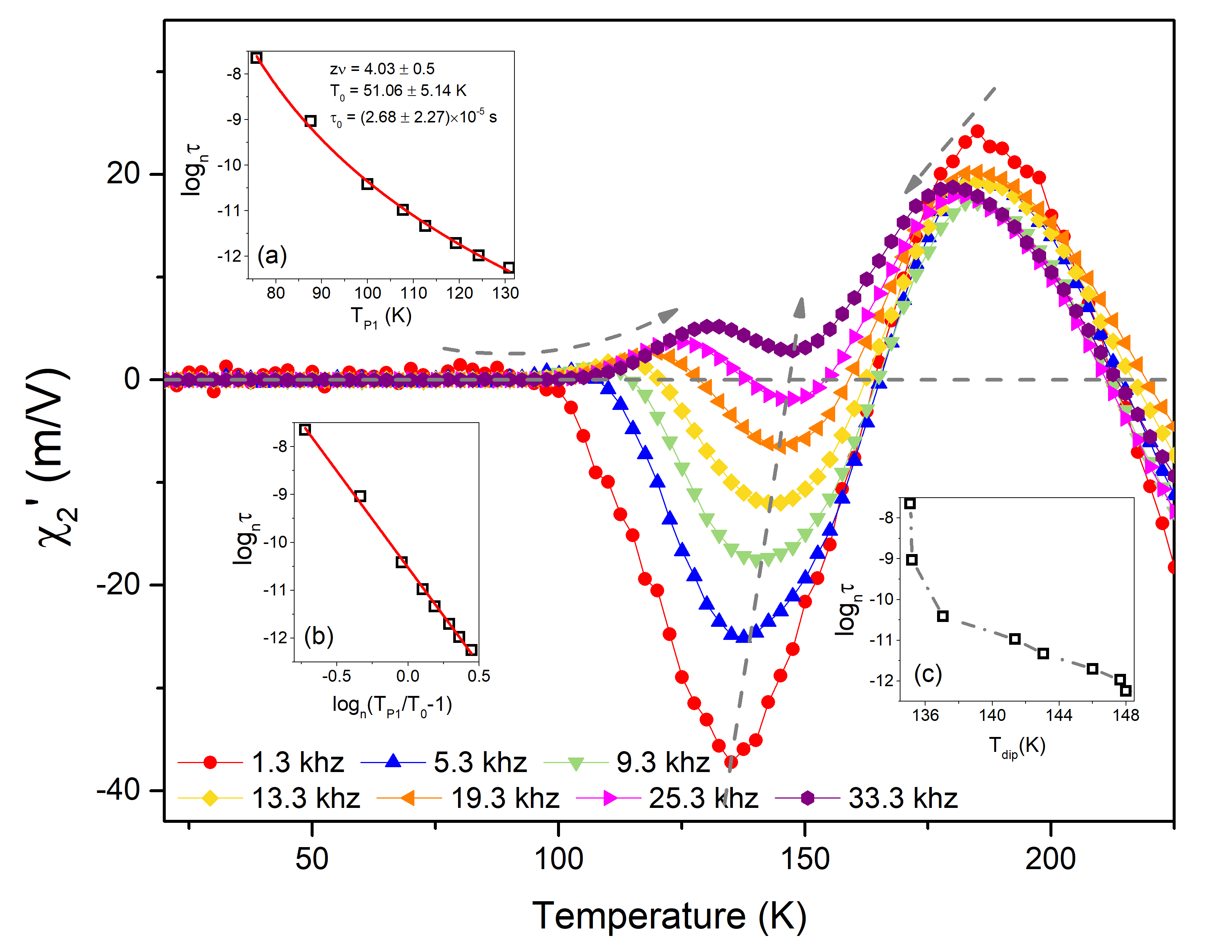}
	\caption{Real second-order susceptibility $\chi_2'$ of Ba$_3$BiRu$_2$O$_9$ as a function of temperature for various frequencies. The dynamics of three observed features: peak $\rightarrow$ dip $\rightarrow$ peak ($T_{P1} \rightarrow T_{dip} \rightarrow T_{P2}$) shown with grey, dashed arrows. Inset (a): $T_{P1}$ shows critical slowing down behaviour, non-linear fit with parameters given. Inset (b): Examining fit quality with a linear log-log plot using $T_0$ obtained from the nonlinear fit. Inset (c): Relaxation process across $T_{dip}$, dotted lines lines are only guide to the eye (not a fit)}
	\label{babi2} 
\end{figure}

This frequency dependence is manifested distinctly in the $\chi_2'$ curves shown in Fig. \ref{babi2} where one can see three different types of dispersion peaks. The first one occuring $\sim$100K, when the PNRs start to develop as the system is warmed up, and the net polarization rises in the system and attains a positive maxima at a temperature ($T_{P1}$) that increases with applied frequency. The dynamic response of these PNRs is seen to follow the critical slowing down behaviour (insets (a) and (b) of Fig. \ref{babi2}) when we plot $T_{P1}$ vs $\tau = \frac{1}{2\pi f}$  The fit parameters obtained are: the dynamical exponent $z\nu$ = 4.03 $\pm$ 0.5 ; the ideal freezing temperature $T_0$ = (51.06 $\pm$ 5.14) K; and the characteristic relaxation time $\tau_0$ = (2.68 $\pm$ 2.27)$\times$10$^{-5}$ s. The value of $\tau_0$ is about 2 orders larger than that reported previously for the temperature regime above the phase transition. This seems to indicate that as the temperature is increased the size of the correlated domains decreases. The second dispersion is seen in the negative Polarization peaks ($T_{dip}$), but unfortunately it does not seem to follow any of the relaxation models. However, the reduction in the amplitude of the negative peak with increasing frequency seems to once again indicate that larger polar domains are associated with a negative polarization while smaller ones are positively aligned with respect to the applied electric field orientation. This type of $\chi_2'$ response is similar to that seen in relaxor FEs like PbMg$_{1/3}$Nb$_{2/3}$O$_3$ (PMN) and Sr$_{0.61}$Ba$_{0.39}$Nb$_2$O$_6$ (SBN61) \cite{miga11}.  The third dispersion is seen in the positive maxima following the negative to positive $P$ crossover for all frequencies, which indicates development of a significant positive polarization in the system.  The frequency dependence of these peaks is reversed (similar to those seen in Ba$_3$CoRu$_2$O$_9$) and very small. It almost seems as if a convergence of two types of relaxation processes occurs at the crossover point. 

\begin{figure}[ht]
	\includegraphics[width=\columnwidth, trim=0.3cm 0.1cm 0.2cm 0.2cm]{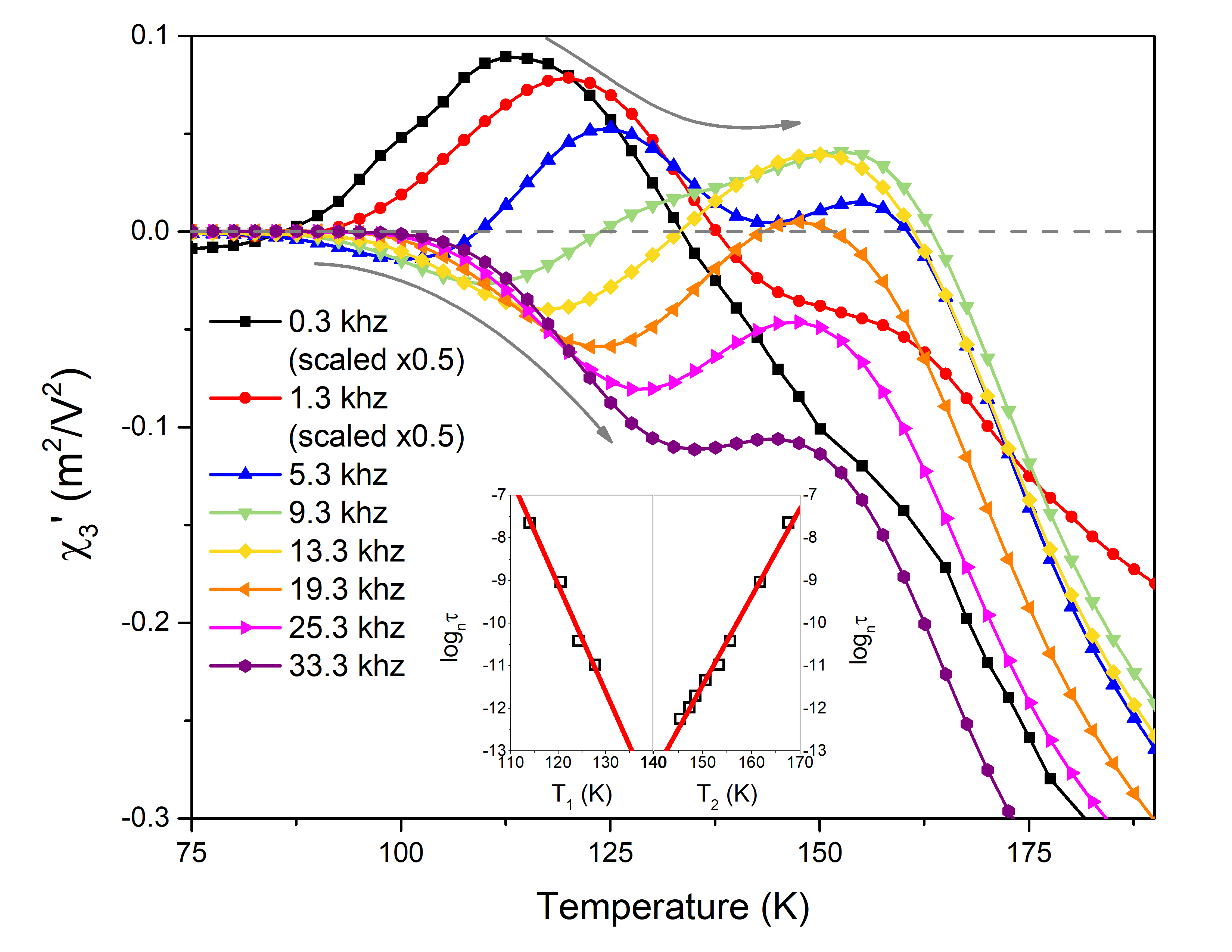}
	\caption{Real third-order susceptibility $\chi_3'$ of Ba$_3$BiRu$_2$O$_9$ as a function of temperature for various frequencies in the temperature range 75-190 K. The coexistence of two types of relaxation processes is captured in dynamics of the two features observed. Inset shows $\tau$ plotted as a function of T$_{1}$ and T$_{2}$. }
	\label{babi3} 
\end{figure}

Similar implication is obtained from $\chi_3'$ as can be seen in Fig. \ref{babi3} where the two distinct anomalies can be seen moving with temperature and frequency.  At the lowest frequency (333 Hz), a single broad peak is recorded below 125 K along with a subtle slope change close to 175 K. Next, at 1.33 kHz, the peak shifts to a slightly higher temperature, but the slope change near 175 K becomes more pronounced and moves to the left. At 5.33 kHz, the hump crosses over to positive values and becomes nearly comparable in amplitude to the broad peak around 125 K. This is the frequency where both peaks can be seen at their best resolution. For successive frequencies, the two peaks continue to come closer until only one of them survives at $\sim$ 150 K. This seems to be a clear indication of multiple relaxation processes acting upon the polar domains within the spin-gap phase which is why characterizing them is nontrivial. This dynamical behaviour is represented in the inset of Fig. \ref{babi3} where log$\tau$ is plotted as a function of both T$_{1}$ and T$_{2}$. It is clear that the relationship is linear and at the highest frequencies the two peaks merge somewhere between 130 and 150K. 

\begin{figure}[ht]
	\includegraphics[width=0.95\columnwidth, trim=0.6cm 0.1cm 0 0.1cm]{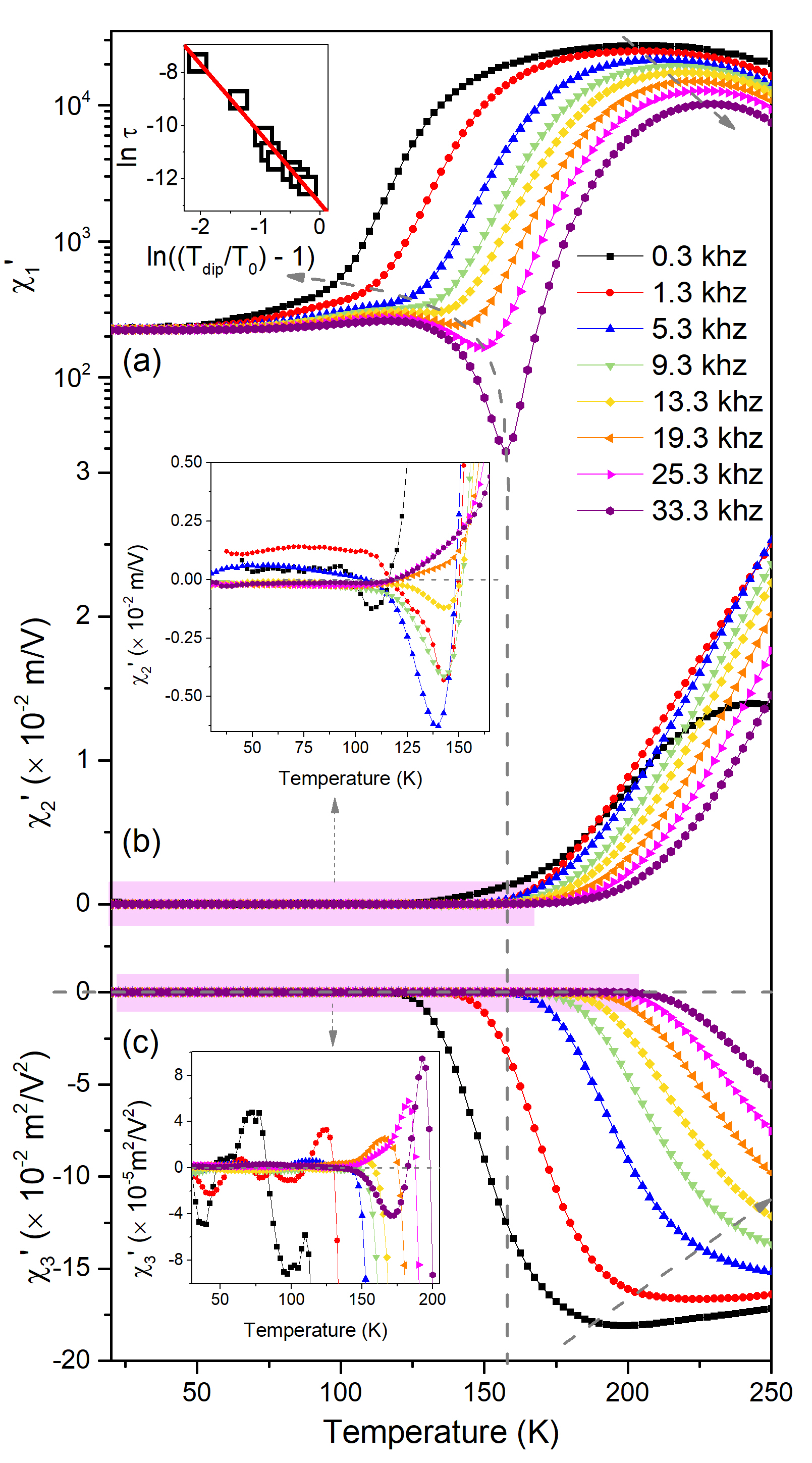}
	\caption{(a) Real first-order susceptibility $\chi_1'$ of Sr$_3$CaRu$_2$O$_9$ as a function of temperature for various frequencies. The sharp divergent feature at 150 K corresponds to the magnetic transition at $\sim$ 160 K. (b) Real second-order susceptibility $\chi_2'$ of Sr$_3$CaRu$_2$O$_9$ as a function of temperature for various frequencies. Inset: Zoom-in of the region below 150 K, highlighting the small frequency dependent negative polarization developed before rising sharply to values an order of magnitude higher. (c) Real part of third-order susceptibility, $\chi_3'$ of Sr$_3$CaRu$_2$O$_9$ as a function of temperature measured for various frequencies. Inset: Zoom-in of the region below 150 K, highlighting the miniscule magnitude of $\chi_3$ maxima on positive scale.}
	\label{fig8} 
\end{figure}

\begin{figure}[ht]
	\includegraphics[width=\columnwidth, trim=1.5cm 0.1cm 0.05cm 0.1cm]{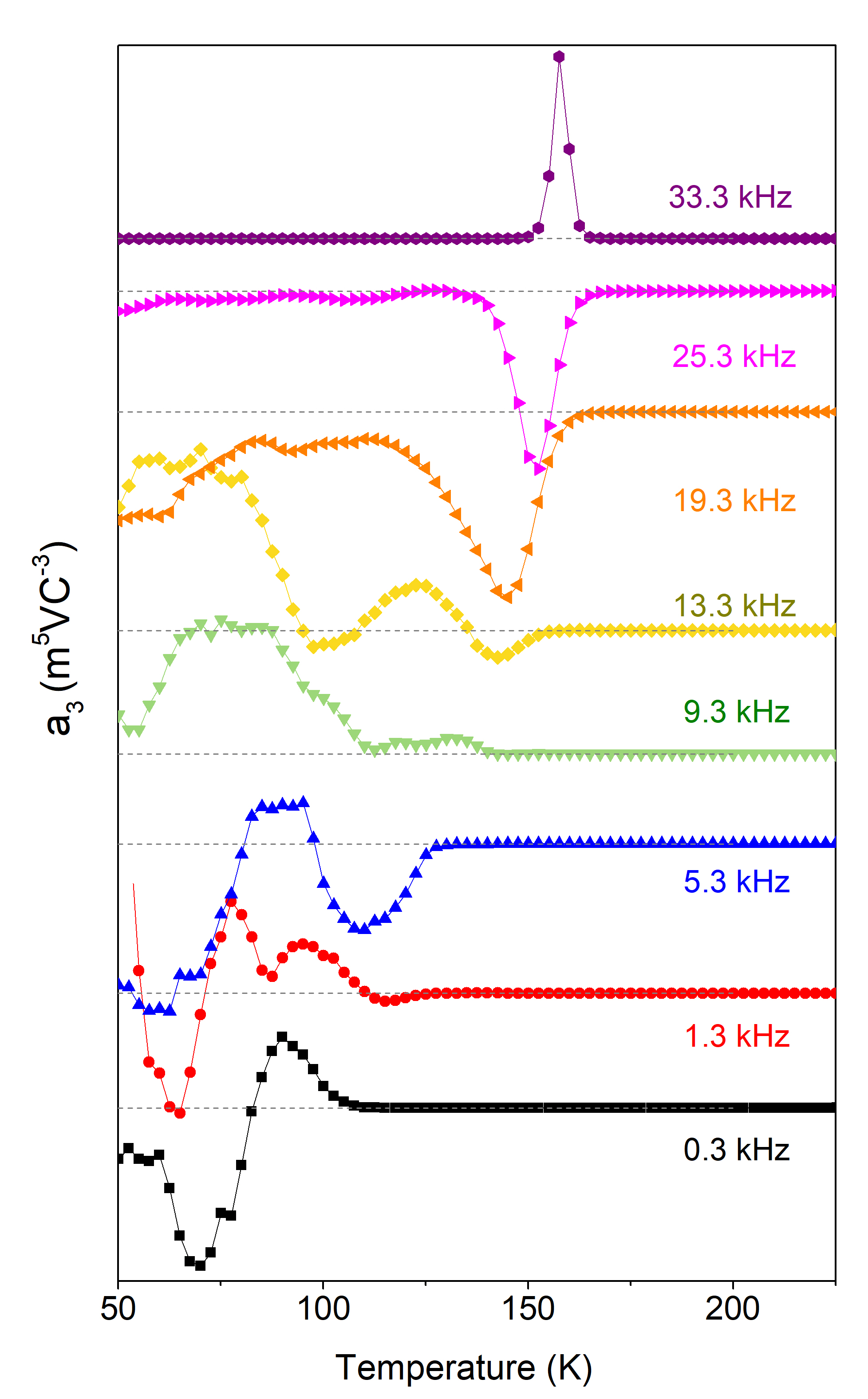}
	\caption{Scaled nonlinear susceptibility, $a_3$ of Sr$_3$CaRu$_2$O$_9$ as a function of temperature shown for frequencies 0.3, 1.3, 5.3, 9.3, 13.3 19.3, 25.3 and 33.3 kHz. Horizontal dotted lines show the position of zero value with respect to each curve.}
	\label{srcaa3} 
\end{figure}

Our final system of interest, Sr$_3$CaRu$_2$O$_9$, is a relatively unexplored compound. Measurements of its linear dielectric permittivity ($\epsilon$)  have only recently been reported \cite{kumar2023} and present the possibility of magnetodielectric behaviour. Fig. \ref{fig8} (a) shows the first-order susceptibility $\chi_1'$ plotted on a log scale, exhibiting a sharp dispersion-like feature at 160 K which is suppressed at lower frequencies. This behaviour is quite similar to that of Ba$_3$BiRu$_2$O$_9$ (Fig. \ref{fig5} (a)) albeit more pronounced at the transition in this case. The relaxation in this regime is described by the critical slowing down behaviour and yields the parameter values: $T_0$ = (87.9 $\pm$ 5.1) K, $\tau_0 \approx$ (2.4 $\pm$ 0.5)$\times$10$^{-6}$s and z$\nu\approx$ 2.6 $\pm$ 0.5. A dispersion can also be seen in the broad peaks close room temperature although the colossal susceptibility values and as well as the dispersion are both likely to be an extrinsic effect of strong surface contribution to polarization resulting from delocalized surface charges at high temperature. Since these materials are semiconducting in nature and not insulating dielectrics, the contribution from leakage current close to room temperatures becomes inevitable. This is despite the fact that Sr$_3$CaRu$_2$O$_9$ has a room temperature resistivity of $\sim$ 0.9 $\ohm$-m which is higher than the other two compounds in this study [$\rho$(Co,300K) = 7$\times$10$^{-4}$ $\ohm$-m, $\rho$(Bi,300K) = 0.3 $\ohm$-m]. This is likely the reason why the lossy behaviour kicks in at the lowest temperatures for the most conducting sample (Ba$_3$CoRu$_2$O$_9$).

As can be seen from Fig. \ref{fig8} (b), the polarization drops dramatically as the system is cooled and enters the magnetically ordered phase below $\sim$200 K. The inset shows the very small crossover to negative values that occurs close to 150 K for the lowest frequencies. The small negative cusp is strongly suppressed with increasing frequency and vanishes beyond 13.3 kHz. For higher frequencies $P$ simply vanishes below 125 K. This is expected since the global crystal symmetry is $P2_1/c$ which is centrosymmetric. The persistence of the small negative polarization for lower frequencies is likely to be a manifestation of cluster-like dipolar segments with corresponding to longer response times. These segments take longer to equilibriate to the ideal zero polarization state leading to the small negative extrema. 

Fig. \ref{fig8} (c) shows $\chi_3'$ as a function of temperature for Sr$_3$CaRu$_2$O$_9$. As can be seen from the inset, $\chi_3'$ for all the frequencies shows a crossover from a small, nearly negligible positive value to massive negative values (almost 4-5 decades larger). The crossover point also shows strong frequency dependence. The rise of this large third-order polarization is gradual, spanning the temperature range 100-200 K. Beyond 200 K, the growth slows down, eventually starting to drop close to room temperature. The dominantly negative value of $\chi_3'$ above the magnetic transition is in agreement with the expectation of SRBRF model in for relaxor FEs. Similar behaviour of $\chi_3'$ has been reported for La$_2$NiMnO$_6$, a partially disordered double perovskite manganite and Ca$_3$Mn$_2$O$_7$, a layered perovskite exhibiting hybrid-improper ferroelectricity (HIF) \cite{sahlot2020}.

Fig. \ref{srcaa3} shows the variation in the scaled nonlinear susceptibility $a_3$ for frequencies 0.3-33.3 kHz. With increasing frequency a negative peak can be seen to develop close to 150 K (corresponding to the positive peaks of $\chi_3'$ just before the crossover to negative values). For the highest frequency however, the peak unexpectedly shows sign reversal accompanied by a strong divergence. It is also of note, that $a_3$ rapidly vanishes beyond the divergence point. As mentioned earlier, $a_3$ is expected to show a critical singularity at $T_f$ in case of dipolar glasses \cite{pirc1994}. So, it seems that Sr$_3$CaRu$_2$O$_9$ contains manifestations of both dipolar glasses and relaxor FEs at the magnetic phase transition. It will be interesting to continue to study this material to discern which one of these is the true/dominant state which goes on to define the low-temperature dipolar ground state in this system.

\section{Conclusions}
In conclusion, we have performed the first extensive nonlinear dielectric measurements on polycrystalline samples of three triple perovskite ruthenates: Ba$_3$CoRu$_2$O$_9$. Ba$_3$BiRu$_2$O$_9$ and Sr$_3$CaRu$_2$O$_9$ and have attempted to study the dynamic behaviour of the underlying polar orders in these systems by utilizing the higher harmonics of the electric polarization response. 
In Ba$_3$CoRu$_2$O$_9$, the magnetoelastic transition at $\sim$100K is captured in $\chi_{1,2,3}$, and the hypersusceptibilities ($\chi_{2,3}$) and $a_3$ shed light on frequency-dependent dynamics occuring just above the transition, which have no analogous feature in the linear susceptibility.
In Ba$_3$BiRu$_2$O$_9$, $\chi_{2,3}$ unveil a dipolar regime lying within the spin-gap phase where multiple relaxation processes seem to coexist, of which some do not conform to the relaxation models routinely used to describe the dynamics of disordered systems. 
In Sr$_3$CaRu$_2$O$_9$, anomalies are observed in the vicinity of the magnetic phase transitions (160 K, 190 K) in $\chi_{1,2,3}$ along with strong dispersion effects. Features in $\chi_3$ and $a_3$ indicate possibility of both dipolar-glass and relaxor FE behaviour and further nonlinear measurements, possibly on single crystals, would be needed to unambiguously ascertain the true behaviour of polar orders in this system. Also, the concurrence with the SRBRF model found in $\chi_3'$ for both Sr$_3$CaRu$_2$O$_9$ and Ba$_3$BiRu$_2$O$_9$ suggests that materials whose polar orders are strongly correlated with spin, lattice or orbital orders provide an alternate class of systems (other than classic/relaxor FEs) to probe concomitant phase transitions and can potentially be exploited as testbeds for theoretical models in the future. 

\begin{acknowledgments}
K. Dey, S. Thadeti and H. Meshram are acknowledged for their help in sample synthesis. S.C. and S.N. acknowledge support from an Air Force Research Laboratory grant (Grant No. FA2386-21-1-4051). SC is grateful to Avirup De for academic discussion and support during instrumentation.  
\end{acknowledgments}

\bibliography{NLDchibib}

\end{document}